\documentclass[preprint,12pt,authoryear]{elsarticle}

\usepackage{amssymb}

\journal{New Astronomy Reviews}

\newcommand\nat{Nature}
\newcommand\apj{ApJ}
\newcommand\apjl{ApJL}
\newcommand\apjs{ApJS}
\newcommand\aj{AJ}
\newcommand\actaa{Acta Astron.}
\newcommand\mnras{MNRAS}
\newcommand\aap{A\&A}
\newcommand\pasp{PASP}

\newcommand\blender{{\tt BLENDER}}
\newcommand\kepler{{\it Kepler}}
\newcommand\ktwo{{\it K2}}

\newcommand\earth{\oplus}
\newcommand\sun{{\odot}}
\newcommand\kms{\ifmmode{\rm km\thinspace s^{-1}}\else km\thinspace s$^{-1}$\fi}
\newcommand\ms{\ifmmode{\rm m\thinspace s^{-1}}\else m\thinspace s$^{-1}$\fi}
\newcommand\cms{\ifmmode{\rm cm\thinspace s^{-1}}\else cm\thinspace s$^{-1}$\fi}

\begin{document}

\begin{frontmatter}



\title{Discovery of the first Earth-sized planets orbiting a star other than our Sun in the Kepler-20 system}


\author[lab1]{Guillermo Torres}
\author[lab1,lab2]{Fran\c cois Fressin}

\address[lab1]{Center for Astrophysics \textbar\ Harvard \&
  Smithsonian, 60 Garden St., Cambridge, MA 02138, USA,
  e-mail: gtorres@cfa.harvard.edu}

\address[lab2]{CVS Health, 1 CVS Dr., Woonsocket, RI 02895, USA,
  e-mail: francoisfressin@yahoo.fr}

\begin{abstract}
Discovering other worlds the size of our own has been a long-held
dream of astronomers. The transiting planets Kepler-20\,e and
Kepler-20\,f, which belong to a multi-planet system, hold a very
special place among the many groundbreaking discoveries of the
\kepler\/ mission because they finally realized that dream. The radius
of Kepler-20\,f is essentially identical to that of the Earth, while
Kepler-20\,e is even smaller (0.87~$R_{\earth}$), and was the first
exoplanet to earn that distinction.  Their masses, however, are too
light to measure with current instrumentation, and this has prevented
their confirmation by the usual Doppler technique that has been used
so successfully to confirm many other larger planets. To persuade
themselves of the planetary nature of these tiny objects, astronomers
employed instead a statistical technique to ``validate'' them, showing
that the likelihood they are planets is orders of magnitude larger
than a false positive. Kepler-20\,e and 20\,f orbit their Sun-like
star every 6.1 and 19.6 days, respectively, and are most likely of
rocky composition.  Here we review the history of how they were found,
and present an overview of the methodology that was used to validate
them.

\end{abstract}


\end{frontmatter}


\section{Introduction}
\label{sec:introduction}

Thanks to the \kepler\ mission we now know that small planets similar
in size to the Earth are common throughout the Galaxy
\citep{Howard:2012, Fressin:2013, Dressing:2013, Dressing:2015,
  Petigura:2013, Marcy:2014, Burke:2015}. What seems so clear now was
not at all obvious at the time the spacecraft was launched in March of
2009, as no such planets had been found outside the solar system.  The
ones discovered until then by the transit method were all Neptune-size
($\sim$4\,$R_{\earth}$) or larger\footnote{The era of smaller planet
  discoveries began in earnest later that same year with the CoRoT
  mission \citep{Rouan:1998, Baglin:2006}, and the announcement of the
  planet CoRoT-7\,b \citep{Leger:2009, Queloz:2009}, an object about
  1.7 times larger than the Earth.}. These had all been confirmed by
measuring their dynamical masses through high precision
radial-velocity observations, to show that they are indeed in the
planetary range. Here we recount the developments that led to the
discovery of the first two Earth-sized exoplanets, Kepler-20\,e
and Kepler-20\,f \citep{Fressin:2012}, which marked a very important
milestone in the field of exoplanet research.  Unlike their larger
cousins that are amenable to Doppler studies, the masses of
Kepler-20\,e and Kepler-20\,f have \emph{not} been measured because
the reflex motion they induce on the host star is too small to
detect. For this reason these objects required the use of an entirely
different analysis technique to assess their planetary nature.

The importance of careful vetting of candidates and of confirmation by
the Doppler technique became painfully obvious as soon as ground-based
transit surveys began reporting planetary candidates. It was quickly
found that the vast majority turned out to be false positives of one
kind or another \citep[see, e.g.,][]{Brown:2003}, with estimates of
the false positive rates reaching as high as 90\% or 95\% in some
cases \citep{Konacki:2003, ODonovan:2006a, Latham:2007}. The most
common types of astrophysical false positives, often referred to as
``blends'', are eclipsing binaries that happen to be along the line of
sight, whether physically associated with the target or not. When this
happens, the otherwise deep eclipses of the binary are greatly
attenuated by the target star and made to look so small that they can
be indistinguishable from the transit signals of true planets.  As it
turns out, however, confirming a planet by measuring the reflex motion
of the parent star is not always feasible. For example, the star may
be too faint, it may be rotating too rapidly, or it may be too
chromospherically active to allow the necessary precision in the
radial velocities. Or even if it does lend itself to Doppler studies,
the signal may simply be too small to detect, if the orbital period is
long and/or the planetary mass too small relative to the mass of the
star.

This was precisely the case for Kepler-20\,e and Kepler-20\,f. As
their designations suggest, these were the fourth and fifth
transit-like signals detected in the photometric observations of
Kepler-20 (KIC\,6850504), a $V = 12.5$, mid G-type star in the
constellation Lyra \citep{Gautier:2012}, and one of the many
multi-planet systems (``multis'') that \kepler\/ would
find. For a review of the discovery and implications of multis,
  we refer the reader to the article by \cite{Steffen:2019} in this
  Special Issue. With transit depths each under 100 parts per million
hinting at objects of both small sizes and small masses, the absence
of corresponding radial-velocity signatures was not all that
surprising. As an alternative to dynamical confirmation, an attempt
was made by \cite{Gautier:2012} to ``validate'' these two signals in a
statistical way using a procedure that had become known as
\blender. The idea behind \blender\ is to simulate blend
configurations all the way through to the light curves they are
expected to produce, and to use the shape of the real transits to
discriminate against as many of those blends as possible.  Other
blends may be rejected if additional observations indicate the
intruding objects would have been uncovered. The technique then aims
to show in a quantitative way that the likelihood of the remaining
false positives is much smaller than that of a true planet. This
approach had been used in a few other cases before, but it was
relatively new at the time and was not sufficient to demonstrate the
planetary nature of Kepler-20\,e and 20\,f to a high enough degree of
confidence. Further improvements to \blender\ would be required, as we
describe below, and the procedure did eventually succeed in showing
beyond a reasonable doubt that the two objects are indeed Earth-sized
or smaller planets, with radii of $R_p = 0.87\,R_{\earth}$ and
$1.03\,R_{\earth}$, respectively, as measured initially by
\cite{Fressin:2012}.

The development of the validation methodology represents a significant
advance in our ability to discover small transiting planets. So far it
is the only alternative we have when the mass cannot be measured
directly, either by the Doppler effect or by modeling transit timing
variations (TTVs) in multi-planet systems. In fact, statistical
validation is now the approach that has verified the largest number of
transiting planets from \kepler\/ and its successor mission \ktwo, and
promises to be invaluable for future space-based transit searches as
well. Because of its considerable impact for small planets
discoveries, and in the spirit of this special issue, we chronicle in
the next section the history of how validation came about, leading up
to its application to Kepler-20\,e and 20\,f.  The more technical
details of the method may be found in the sources cited below.

\section{Statistical validation: a pathway to the discovery of small planets}
\label{sec:validation}

Readers familiar with the early history of photometric searches for
transiting planets may recall that the very first lists of candidates,
following the momentous discovery of HD\,209458\,b
\citep{Charbonneau:2000, Henry:2000}, were released by the Optical
Gravitational Lensing Experiment \citep[OGLE;][]{Udalski:2002a,
  Udalski:2002b}. Out of one of those lists emerged the second known
transiting planet, OGLE-TR-56\,b \citep{Konacki:2003}, a Jovian-size,
Jovian-mass object with an orbital period of just 1.2 days that was
also the first to be discovered in a photometric
survey\footnote{HD\,209458\,b, the first known transiting exoplanet,
  was originally found in a radial-velocity survey and only later
  discovered to undergo transits.}. Although it was confirmed
dynamically, the mass determination for OGLE-TR-56\,b was based on few
and very challenging radial-velocity observations given the faintness
of the host star ($V = 16.6$), and extra precaution was taken to
examine possible false positive scenarios, particularly since the
target is projected against the crowded field of the Galactic center.

It is in this context that one of us (G.T.) developed numerical
procedures to simulate realistic light curves resulting from a blend
with an eclipsing binary along the line of sight. A wide range of
false positive scenarios were generated and compared against the OGLE
photometry, assuming different binary properties and also varying the
relative distance between the binary and the target. These tests
revealed that many of the configurations resulted in light curves that
fit the real observations just as well as a planetary transit model,
matching both the depth and the overall shape of the transits.
However, it was also found that for all of these blends the predicted
brightness of the eclipsing binary together with the radial-velocity
semi-amplitude of its primary component would be expected to cause
noticeable asymmetries in the spectral line profiles, or even the
presence of a second set of lines moving around with the 1.2-day
period of the signal, and yet neither of these was seen. This provided
further support for the planetary nature of the object. The
mathematical details of the technique were laid out by
\cite{Torres:2004}, along with an application to another candidate
that \emph{did} turn out to be a false positive (OGLE-TR-33).

The capability was later added to generate blend light curves
simultaneously in other passbands, and to predict the overall colors
of the blend, both of which can supply further useful checks against
follow-up observations that might be available. Simulations like
these, coupled with an analysis of the spectral line bisectors
\citep[e.g.,][]{Torres:2005}, were used also by other ground-based
transit surveys as additional insurance against blends
\citep{Mandushev:2005, ODonovan:2006a, ODonovan:2006b, Bakos:2007}.
However, in cases where planets were confirmed in these surveys, it
was always a mass measurement via radial velocities that provided the
final proof, and the blend simulations played only a supporting role.

\subsection{First application to small planets: Kepler-9\,d}

 When the \kepler\/ mission began finding very small transit-like
  signals that could not be confirmed dynamically, it became clear
  that the simulation approach would come to be crucial. A way was
  needed not only to improve the ability to reject false positives as
  much as possible, but more importantly, to quantify the probability
  that any one of the remaining blends might actually be causing the
  small drops in brightness. The first real test came with
Kepler-9\,d \citep{Torres:2011}, a super-Earth with a size of about
1.6\,$R_{\earth}$.  This was the third signal found in the
multi-planet system Kepler-9 \citep{Holman:2010} featuring two larger
Saturn-sized objects ($\sim$9\,$R_{\earth}$), which were also the
first to display unambiguous TTVs \citep[see the review article
    by][this Special Issue]{Ragozzine:2019}.

The simulations for Kepler-9\,d were expanded to include a more
complete, grid-based exploration of parameter space for false
positives. In addition to eclipsing binaries, the numerical
experiments now included scenarios involving an intruding single star
located anywhere along the line of sight that is transited by a larger
planet, rather than by another star. While it may be argued that this
type of contaminant is in itself a bona fide planetary system, the
unfortunate alignment with the star one is interested in causes the
transits to appear shallower, simulating the presence of a smaller
planet orbiting the target star. As the goal was to prove the
existence of a planet of small size around the target (rather than a
larger one of unknown size around the companion), these configurations
were considered as false positives. Allowance was made also for
eccentric orbits for all categories of simulated blends, and for
differential interstellar extinction between the intruding star or
binary and the target. Additionally, a more thorough use was made of
available follow-up observations to help rule out blends. Detection
limits from high-resolution imaging were now taken into account, as
well as limits on unseen spectroscopic companions in high-resolution
spectra, measured color indices for the target, and other limits on
nearby companions based on an analysis of the flux centroids from the
\kepler\ images themselves. All of these helped, but many false
positive scenarios still remained viable.

The expected numbers of viable false positives of different kinds is
of course a function of the number density of stars at the sky
location of the target, and depends also on how common eclipsing
binaries and larger planets are.  For Kepler-9\,d these blend
frequencies were calculated in discrete magnitude bins by counting up
the ones that were permitted by all observational constraints, using
number densities from Galactic structure models along with estimates
of the rates of occurrence of eclipsing binaries and larger planets
from the early \kepler\/ results.  It was also realized that in order
to obtain a proper false alarm probability (FAP), or equivalently a
confidence level that the signal is due to a true super-Earth-sized
planet, an estimate was required of the rate of occurrence of such
planets. Expressed in terms of the numbers of expected false positives
and planets, ${\rm FAP} = N_{\rm FP}/(N_{\rm FP} + N_p)$. However, the
planet occurrence rate (referred to as the ``planet prior'') was not
well known at the time, so arguments were made drawing on statistics
from Doppler surveys, on theoretical considerations, and on
preliminary \kepler\/ results that were based on candidate detections
rather than confirmed planets. The most conservative of those
estimates allowed Kepler-9\,d to be validated to a sufficiently high
level of confidence corresponding to a false alarm probability of $6
\times 10^{-4}$ \citep{Torres:2011}.

This framework for simulating blend scenarios and performing
statistical validation became known as \blender, and over the next
year or so it was applied in a few other cases with relatively minor
changes. The software to perform the computationally intensive blend
simulations and map out parameter space was ported to the Pleiades
supercomputer at the NASA Ames Research Center (California, USA), with
the help of Chris Henze.

\subsection{The challenge of even smaller planets}

Kepler-20\,e and 20\,f were more demanding still than Kepler-9\,d
because of the shallower transits and the associated smaller
signal-to-noise ratios. This meant they contain less information with
which to constrain the detailed shape of the transit and rule out
blends. The simple-minded procedure of tabulating blends in discrete
magnitude bins to compute their frequencies was replaced by a more
sophisticated Monte Carlo approach. Background stars were drawn at
random from a Galactic structure model, and were assigned either a
stellar or a planetary companion, depending on the type of false
positive. This was done taking into account the known properties of
eclipsing binaries and the size distribution of larger planet
candidates, as determined from the \kepler\/ mission itself. For
blends consisting of a planet transiting another star physically
associated with the target (hierarchical triple configuration) the
simulations placed such companions in randomly oriented orbits around
the host star following the known distributions of periods, mass
ratios, and eccentricities of binary systems. The frequencies for each
type of false positive were then calculated after removing
configurations inconsistent with constraints from the lightcurve
morphology and the follow-up observations. The outcome of this
exercise for Kepler-20\,e gave a blend frequency of background
eclipsing binaries of $3.1 \times 10^{-8}$, a frequency of background
stars transited by larger planets of $2.1 \times 10^{-7}$, and a
frequency of hierarchical triple configurations of $5.0 \times
10^{-7}$. These three contributions added up to a total blend
frequency of $7.4 \times 10^{-7}$. For Kepler-20\,f the numbers were
$1.2 \times 10^{-6} + 4.5 \times 10^{-7} + 3.6 \times 10^{-6} \approx
5.2 \times 10^{-6}$. The planet priors, i.e., the a priori chance that
the parent star Kepler-20 has a planet of a similar size as implied by
each of the two signals, were estimated again using the catalog of
\kepler\/ objects of interest (KOIs), which had expanded by then. As
KOIs are still only candidates, the conservative assumption was made
that only 10\% of them are real planets, even though other estimates
at the time were nearly an order of magnitude larger
\citep[e.g.,][]{Morton:2011}. The resulting planet priors were $3.1
\times 10^{-4}$ for Kepler-20\,e and $7.5 \times 10^{-4}$ for
Kepler-20\,f.

By this time it had already been shown that multi-planet systems such
as Kepler-20 tend to be coplanar \citep{Lissauer:2011}. Because
Kepler-20 was already known to have three other transiting planets
(Kepler-20\,b, 20\,c, and 20\,d; see below), this made it much more
likely that Kepler-20 has a transiting planet at the periods of
Kepler-20\,e and 20\,f than a random \kepler\ target. With this
``multiplicity boost'', the planet priors for 20\,d and 20\,f
increased to $2.5 \times 10^{-3}$ and $7.1 \times 10^{-3}$,
respectively. Comparing these to the total blend frequencies from
above, the false alarm rates became $3.0 \times 10^{-4}$ for
Kepler-20\,e and $7.3 \times 10^{-4}$ for Kepler-20\,f, which were
deemed sufficiently small to declare the candidates validated as true
Earth-sized planets.

Beyond the success in demonstrating the planetary nature of the first
two known Earth-sized planets, statistical validation using
\blender\ has been applied to many other planets, including some of
the most iconic discoveries of the \kepler\ mission. Examples include
\kepler's first rocky planet \citep[Kepler-10\,b;][]{Batalha:2011},
the first small planets in the habitable zone of their parent
  stars (Kepler-22\,b, \citealt{Borucki:2012}; and Kepler-62\,f,
  \citealt{Borucki:2013}; see also the article by
  \citealt{Borucki:2019} in this Special Issue), the first two
transiting planets ever discovered in a cluster \citep[Kepler-66\,b
  and Kepler-67\,b;][]{Meibom:2013}, the discovery of a
sub-Mercury-sized planet \citep[Kepler-37\,b;][]{Barclay:2013}, a
transiting planet near the snow line of its parent star
\citep[Kepler-421\,b;][]{Kipping:2014}, a super-Earth in the habitable
zone of a G2 star with an orbital period near one year
\citep[385~days, Kepler-452\,b;][]{Jenkins:2015}, the discovery of a
sub-Neptune-sized planet in the open cluster Ruprecht~147
\citep[K2-231\,b;][]{Curtis:2018}, and others.

Statistical validation as an exoplanet discovery tool when Doppler
confirmation is not feasible is now mainstream, and while
\blender\ led the way, several other versions of the same approach
with different strengths have now been implemented that were inspired
by \blender, such as {\tt vespa} \citep{Morton:2012} and {\tt PASTIS}
\citep{Diaz:2014}. These methods all work by comparing priors
  from various scenarios (true planets and false positives) to arrive
  at a confidence level for planethood. A different technique to
  validate candidates in multi-planet systems was developed by
  \cite{Lissauer:2012}, and refined by \cite{Lissauer:2014}, which is
  based on planet multiplicity statistics. With reasonable assumptions
  on the nature and distribution of false positives, these authors
  showed that almost all multi-planet candidates are true planets
  rather than false positives, and that the higher the multiplicity,
  the more likely the candidates are real planets. This immediately
  allowed the validation in bulk of hundreds of \kepler\ planets in
  multis.

Interestingly, of the several thousand exoplanets now known, the vast
majority were actually validated (most with {\tt vespa}, or based
  on multiplicity statistics when in multis) rather than confirmed
dynamically \citep[see, e.g.,][]{Rowe:2014, Morton:2016,
  Crossfield:2016, Mayo:2018}. This is partly a reflection of the fact
that small planets with undetectable Doppler signals far outnumber
larger ones, that many of the \kepler\/ host stars are faint and
unsuitable for Doppler studies, and that observing facilities capable
of high-precision (\ms) radial-velocity measurements are still very
few and far between.

\section{Kepler-20\,e and Kepler-20\,f: two planets the size of the Earth}
\label{sec:kepler-20}

The discovery of the Kepler-20 multi-planet system was announced to
the community by \cite{Gautier:2012}. It featured three transiting
planets (20\,b, 20\,c, 20\,d) with orbital periods ranging from 3.7 to
78~days that were confirmed in the traditional way, and that have
sizes estimated by those authors of 1.9, 3.1, and 2.8\,$R_{\earth}$.
The detections were based on eight quarters of \kepler\/ long-cadence
(30 min) observations made between 2009 May and 2011 March. The same
paper gave news of the detection of two additional transit-like
signals in the same star that were much shallower and had periods of
6.1 and 19.6 days, respectively, but they were left unconfirmed, as
mentioned earlier.  The validation of these two signals as
Kepler-20\,e and 20\,f, the first two Earth-sized planets, was left to
\cite{Fressin:2012}. The planetary sizes reported by these authors,
based on a determination of the properties of the host star and fits
to the light curves, were $0.868_{-0.096}^{+0.074}\,R_{\earth}$ and
$1.03_{-0.13}^{+0.10}\,R_{\earth}$, respectively. \cite{Fressin:2012}
noted also that the first of these planets, Kepler-20\,e, is
potentially smaller than Venus. Until then the smallest known
exoplanet around a Sun-like star had been Kepler-10\,b, with a measured radius of
1.42\,$R_{\earth}$ \citep{Batalha:2011}. While the difference may not
seem all that significant, the validation of Kepler-20\,e and 20\,f
was seen by many as crossing a threshold of sorts, advancing the
frontier of discovery into the realm of planets the size of our own,
and smaller.

The Kepler-20 multi-planet system has received additional attention
more recently. \cite{Buchhave:2016} reported new radial-velocity
measurements, and revisited both the stellar parameter determination
and the photometric solutions, now using the full complement of
short-cadence (1 min) observations of the star obtained in Quarters 3
through 17, rather than the smaller number of long-cadence data used
previously. This is important because the very brief ingress and
egress phases of the transit that are so critical for constraining the
planetary sizes are much better resolved with the 1 min integrations than
the 30 min integrations. The new stellar properties also benefited from
asteroseismic constraints the authors were able to extract from the
short-cadence observations.

\begin{table}[!t]
\caption{Main properties of Kepler-20\,e and 20\,f
  \citep{Buchhave:2016} \label{tab:20ef}}
\vspace{2pt}
\centering
\begin{tabular}{l c c}
\hline\hline \\ [-2.5ex]
Parameter & Kepler-20\,e & Kepler-20\,f \\
\hline \\ [-2ex]
$P$ (days)           & $6.098523_{-0.000014}^{+0.000006}$ & $19.57758_{-0.00012}^{+0.00009}$ \\ [0.5ex]
$T_c$*               & $968.9315_{-0.0007}^{+0.0022}$     & $968.2071_{-0.0043}^{+0.0061}$   \\ [0.5ex]
$R_p$ ($R_{\earth}$) & $0.865_{-0.028}^{+0.026}$          & $1.003_{-0.089}^{+0.050}$        \\ [0.5ex]
$i$ (deg)            & $87.6_{-0.1}^{+1.1}$               & $88.79_{-0.07}^{+0.43}$          \\ [0.5ex]
$a$ (au)             & $0.0639_{-0.0014}^{+0.0019}$       & $0.1396_{-0.0035}^{+0.0036}$     \\ [0.2ex]
$T_{\rm eq}$ (K)**   & $1040 \pm 22$                      & $705 \pm 16$                     \\ [0.2ex]
\hline \\ [-2.5ex]
\multicolumn{3}{l}{*~~Time of mid transit expressed as BJD$- 2,\!454,\!000$.} \\
\multicolumn{3}{l}{**~Equilibrium temperatures taken from \cite{Fressin:2012}.} \\
\end{tabular}
\end{table}

The updated properties obtained by \cite{Buchhave:2016} for
Kepler-20\,e and 20\,f are listed in Table~\ref{tab:20ef}. They
include the orbital semimajor axes and the equilibrium temperatures,
on the assumption of full energy redistribution and a Bond albedo of
0.3. The orbits were assumed to be circular. The planetary sizes are
considerably better determined than before, although the actual values
differ little from those of \cite{Fressin:2012}. This is the result of
a trade-off in the recent work between a small increase in the stellar
radius and a small reduction in the radius ratios
$R_p/R_{\star}$. Further improvements in the stellar properties of
Kepler-20 are now possible thanks to the availability of a precise
parallax from the second data release (DR2) of the {\it Gaia} catalog
\citep{Gaia:2018}, which places the star at a distance of $282.5 \pm
1.7$~pc. Using that measurement and a different methodology,
\cite{Berger:2018} have reported a stellar radius of $R_{\star} =
0.887_{-0.035}^{+0.037}\,R_{\sun}$, which is about 8\% smaller than
the determination by \cite{Buchhave:2016} of $0.964 \pm
0.018\,R_{\sun}$ (a 1.9$\sigma$ difference). The new value would
reduce the sizes of Kepler-20\,e and 20\,f even further to about 0.80
and 0.92\,$R_{\earth}$, respectively, making them both nominally
smaller than the Earth, though with uncertainties increased by a
factor of two.

\begin{table}[!b]
\caption{Main properties of the larger Kepler-20 planets
  \citep{Buchhave:2016} \label{tab:20bcdg}}
\vspace{2pt}
\centering
\begin{tabular}{l c c c c}
\hline\hline \\ [-2.5ex]
Parameter & Kepler-20\,b & Kepler-20\,c & Kepler-20\,g & Kepler-20\,d \\
\hline \\ [-2.5ex]
$P$ (days)              & 3.696     & 10.854    & 34.940    & 77.611   \\
$T_c$*                  & 967.50201 & 971.60796 & 967.50027 & 997.7303 \\
$R_p$ ($R_{\earth}$)    &   1.868   &   3.047   & \ldots    &   2.744  \\
$i$ (deg)               &   87.4    &    89.8   & $<88.7$** &   89.7   \\
$a$ (au)                &  0.0463   &  0.0949   & 0.2055    &  0.3506  \\
$e$                     &   0.03    &    0.16   &  0.15     & \ldots   \\
$K$ (\ms)               &   4.20    &    3.84   &  4.10     &   1.57   \\
$M_p$ ($M_{\earth}$)*** &   9.7     &   12.8    &  20.0     &   10.1   \\
$\rho_p$ (g cm$^{-3}$)  &   8.2     &    2.5    & \ldots    &   2.7    \\
$T_{\rm eq}$ (K)        &  1105     &   772     &  524      &   401    \\
\hline \\ [-2.5ex]
\multicolumn{5}{l}{*~~Time of mid transit expressed as BJD$- 2,\!454,\!000$.} \\
\multicolumn{5}{l}{**~Upper bound on inclination inferred here from the lack of transits.} \\
\multicolumn{5}{l}{***For Kepler-20\,g this is the minimum mass $M_p \sin i$.} \\
\end{tabular}
\end{table}

\section{Kepler-20 architecture, formation, and planetary composition}
\label{sec:architecture}

\begin{figure}[!t]
\centering
\includegraphics[width=12cm]{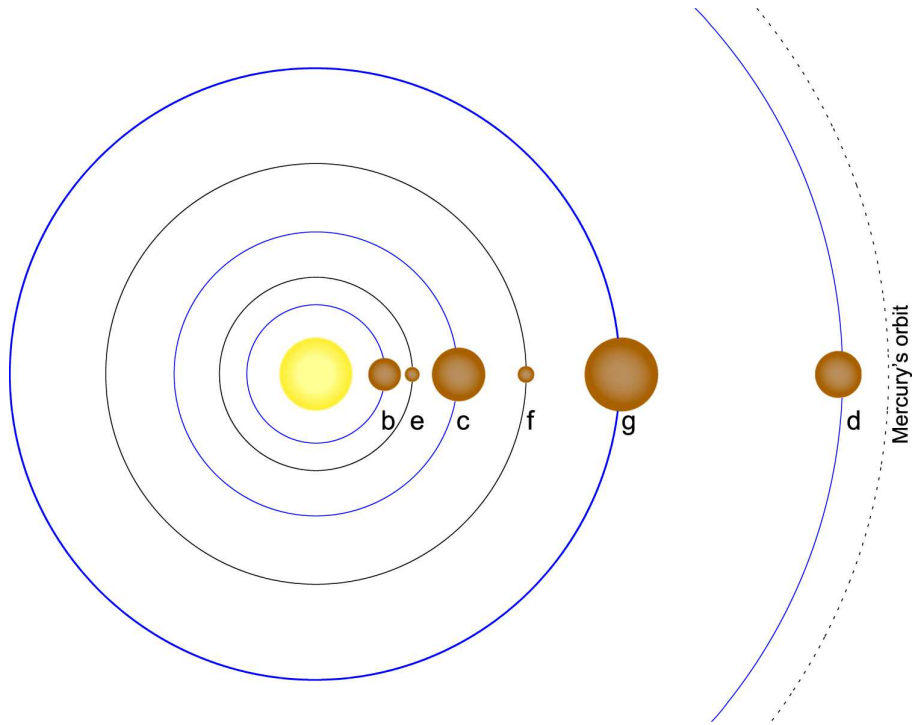}

\caption{{\bf The Kepler-20 system:} Orbital configuration of the
  Kepler-20 system, reproduced from Figure~1 of
  \cite{Buchhave:2016}. All six planets are contained within the
  orbital distance of Mercury in our Solar System. Orbital distances
  are drawn to scale, and planet sizes are rendered in correct
  proportion to each other, though not on the same scale as the
  orbits. The size of Kepler-20\,g was estimated using its mass and
  assuming a composition similar to Kepler-20\,c. Orbits drawn in blue
  represent planets with mass measurements.\label{fig:orbits}}
\end{figure}

With their new radial-velocity measurements \cite{Buchhave:2016} were
able to improve the mass determinations for the three larger planets
Kepler-20\,b, 20\,c, and 20\,d, but the two smaller ones remain below
the detection threshold. Even assuming a rocky composition, which
would maximize their radial-velocity signal, they are expected to
induce reflex motions on the star with semi-amplitudes of only
$\sim$20~\cms.  Interestingly, however, the new Doppler measurements
have revealed a sixth planet (Kepler-20\,g) in this already
extraordinary system, which does not undergo transits. It is nestled
between the outer two previously known planets 20\,f ($P = 19.6$~days)
and 20\,d ($P = 77.6$~days), and revolves around the star once every
34.9~days.  \cite{Buchhave:2016} report it to be more massive than any
of the others, with a minimum mass of $M_p \sin i =
20.0_{-3.6}^{+3.1}\,M_{\earth}$, which is larger than that of Neptune.
For completeness, we summarize in Table~\ref{tab:20bcdg} the main
properties of this new planet and the remaining ones in the system. We
include the measured velocity semi-amplitudes, $K$, the estimated
orbital eccentricities, the planetary masses, and the mean densities,
$\rho_p$.  A schematic view of the architecture of the Kepler-20
system is shown in Figure~\ref{fig:orbits}.

Kepler-20 is quite remarkable in that it is very compact: its six
known planets, five of which transit, are all packed within the
orbital distance of Mercury in our own solar system. Compactness has
now been found to be a feature of many other multi-planet systems as
well. Furthermore, \cite{Gautier:2012} pointed out a striking feature
of Kepler-20 that is the presence of small and likely rocky Earth-size
planets (20\,e and 20\,f) interspersed between larger sub-Neptune
planets at smaller and larger orbital semimajor axes. This is quite
different from our own solar system, in which the terrestrial planets,
gas giants, and ice giants are neatly segregated in regions with
increasing distance from the Sun. Recent studies of samples of
  multi-planet systems have found that the patterns in planet sizes,
  masses, and spacing are linked through formation and subsequent
  orbital dynamics, although the full complexity of planetary system
  architectures is not yet well understood \citep[see,
    e.g.,][]{Millholland:2017, Weiss:2018}.

The long-term stability of the Kepler-20 system was investigated
numerically by \cite{Gautier:2012} prior to the discovery of the
massive non-transiting planet Kepler-20\,g, and by
\cite{Buchhave:2016} afterward. Both studies assumed masses for
Kepler-20\,e and 20\,f of about 0.65 and 1.0~$M_{\earth}$,
respectively, and neither found any indication of instability over a
10 million year period, provided the eccentricities (which are still
poorly determined) are small. \cite{Buchhave:2016} concluded that the
Kepler-20 system is consistent with being dynamically cold, with
relatively small eccentricities and inclination angles, and that it
may have formed during the transition phase when the circumstellar
disk has a high solid surface density but a low or moderate gas
surface density, according to theoretical modeling by \cite{Lee:2016}.

Kepler-20\,e and 20\,f are so small that they most likely have a rocky
composition like the Earth. Based on the properties of the parent star
and the orbital semimajor axes of these two bodies, we estimate they
now receive, respectively, about 187 and 39 times the incident
radiation that the Earth receives from the Sun. Any primordial gaseous
envelopes would have been completely lost to atmospheric
photoevaporation \citep[e.g.,][]{Lopez:2012} or perhaps other
processes such as impact erosion \citep[e.g.,][]{Inamdar:2015}.

Thanks to the improved mass determinations by \cite{Buchhave:2016} for
the other transiting planets in the system, their nature is now also
coming into better focus. Despite the large radius of Kepler-20\,b,
the innermost planet, interior structure models by \cite{Zeng:2013}
indicate it has a rocky composition that is consistent with an
iron-to-silicate ratio similar to that of the Earth. We are likely
seeing the bare core of a planet that lost its primordial atmosphere
due to strong irradiation from the star, equivalent to more than 350
times the flux of the Sun impinging on the Earth. The masses and radii
of Kepler-20\,c and 20\,d, on the other hand, lead to densities that
indicate the presence of volatiles and/or a hydrogen/helium envelope.

\section{Final words}
\label{sec:words}

Kepler-20\,e and 20\,f marked the first time astronomers were able to
verify the presence of another world the size of our own around a
Sun-like star. Since then, efforts have continued to push toward ever
smaller planets, not only from space but also from the ground. As of
this writing there are some 150 transiting planets known that are
about the size of the Earth or smaller\footnote{Count based on the
  tabulation at {\tt http://exoplanets.org/table}, consulted on
  November 15, 2018.}, all but a handful being
\kepler\ discoveries. The exceptions are some of the planets in the
fascinating multi-planet system TRAPPIST-1 \citep{Gillon:2017},
detected recently from the ground around a nearby late-type M dwarf,
and observed also by \ktwo\ as well as {\it Spitzer}. The
record-holder in terms of the smallest measured size is still
Kepler-37\,b \citep{Barclay:2013}, a body that is smaller than the
planet Mercury in our solar system.

In this growing collection of small planets only a few that reside in
multi-planet systems have been confirmed dynamically, not by the
Doppler technique but by taking advantage of their TTVs to measure
their masses. An example is the TTV measurement of the mass of
  the Mars-sized planet Kepler-138\,b by \cite{Jontof-Hutter:2015}.
The rest have all been validated statistically. The use of the
validation approach that has been so successful is likely to continue
into the future in support of missions such as the Transiting
Exoplanet Survey Satellite \citep[TESS;][]{Ricker:2015}, which has now
begun to scrutinize the sky looking for small planets like
Kepler-20\,e and 20\,f around {\bf bright} nearby stars.

\vskip 10pt
\noindent{\bf Acknowledgments}
\vskip 5pt

We thank Jack Lissauer for the invitation to write this review, 
  and for helpful comments.  Our two referees also provided helpful
  suggestions that improved the manuscript. We wish to acknowledge
the invaluable help of Chris Henze (NASA/Ames), who implemented
important modifications to the \blender\ program to improve the
mapping of the range of possible blends, and managed the processing on
the Pleiades supercomputer.  Finally, we are deeply grateful to the
engineers, managers, and scientists who were responsible for the
\kepler\ mission that has led to so many important discoveries
including Kepler-20\,e and 20\,f.

Declarations of interest: none.


\end{document}